\documentclass[aps,prl,amsfonts,amssymb,epsfig,graphics,showpacs,10pt,twocolumn]{revtex4}

\usepackage{graphicx}
\usepackage{epsfig}

\begin{document}

\def\ve{\varepsilon}
\def\cG{{\cal G}}
\def\us{\underline{s}}
\def\expect{{\mathbb E}}
\def\E{{\mathbb E}}
\def\Ball{{\sf B}}
\def\reals{{\mathbb R}}
\def\integers{{\mathbb Z}}
\def\<{\langle}
\def\>{\rangle}
\def\elb{\widehat{\ell}}
\def\de{{\rm d}}
\def\qb{\overline{q}} 
\def\cO{{\cal O}}
\def\xb{{x}}
\def\yb{{y}}
\def\Lb{\widehat{L}}
\def\cL{{\cal L}}
\def\1RSB{\rm\tiny 1RSB}

\date{\today}

\title{Analytic determination of dynamical and mosaic length scales in a Kac glass model}
\author{S.~Franz$^{\,1,3}$ and A.~Montanari$^{\,2,3}$}
\affiliation{$^{1\,}$The Abdus Salam ICTP, Strada Costiera 11, P.O. Box 586, 
I-34100 Trieste, Italy}
\affiliation{$^{2\,}$Laboratoire de Physique Th\'{e}orique de l'Ecole Normale
  Sup\'{e}rieure, CNRS-UMR 8549}
 \affiliation{$^{3\,}$Isaac Newton Institute for Mathematical Sciences
20 Clarkson Road, Cambridge, CB3 0EH, U.K.}


\begin{abstract}

  We consider a disordered spin model with multi-spin interactions
  undergoing a glass transition.
  We introduce a dynamic and
  a static length scales and compute them in the Kac limit
  (long--but--finite range interactions). They diverge at the dynamic
  and static phase transition with exponents (respectively) $-1/4$ and
  $-1$. The two length scales are approximately equal well above the
  mode coupling transition. Their discrepancy increases rapidly as
  this transition is approached. We argue that this signals a
  crossover from mode coupling to activated dynamics.
\end{abstract}

\maketitle

\vspace{-0.6cm}

The relaxation mechanisms of supercooled liquids near the glass
transition are today poorly understood.  For moderate supercooling,
mode coupling theory (MCT) captures several important dynamical
features, such as time scale separation and the relation between
$\alpha$ and $\beta$ relaxation \cite{MCT}.  Unfortunately MCT fails
at low temperature since it predicts a spurious power law divergence of the
relaxation time at an ergodicity breaking temperature $T_{\rm d}$.

A partial elucidation of ergodicity restoration below $T_{\rm d}$
originated from the remark that MCT is exact for some mean-field
disordered spin models \cite{W1}.  In this context, the relaxation
time divergence at $T_{\rm d}$ finds its root in the proliferation of
metastable states with diverging lifetime.  The existence of
such states provides the basic ergodicity restoration force
in finite dimension.  Phenomenological droplet arguments based on this
mean field picture \cite{W2,BB2} predict in fact the existence of
about $\exp\{\Sigma\ell^{d}\}$ different states at length scale
$\ell$, while a given boundary condition can select one of them by
raising the free energy of the others by at most $\Delta F =
\sigma\ell^{\theta}$ with $\theta\le d-1$. Balancing these two forces
yields a ``mosaic length'' $\ell_{\rm
  s}\approx(\beta\sigma/\Sigma)^{\frac{1}{d-\theta}}$, above which the
system is thermodynamically a liquid.  

The above picture suggests that the typical size of ergodicity
restoring rearrangements is determined by thermodynamics.  According
to an alternative intuition
(largely inspired by kinetically constrained models \cite{KCM}), 
as the temperature is lowered, it becomes
impossible to relax local degrees of freedom
through purely local moves. The relevant length scale is essentially
dynamical and corresponds to the size of 
cooperative rearrangements.

In both scenarios, the typical size of rearranging regions can be
characterized by a dynamical (four points) susceptibility and a
corresponding dynamical length $\xi_{\rm d}$ \cite{chi4,BB1}.  Recent
experiments on colloidal systems and supercooled liquids have indeed
revealed a marked increase of $\xi_{\rm d}$ close to the glass
transition \cite{science}. Within MCT, $\xi_{\rm d}$ has been shown to
diverge at $T_{\rm d}$ \cite{chi4,BB1} with no static 
counterpart. 

A relation between mosaic length and the dynamical length $\xi_{\rm d}$ was
first noticed by Jack and Garrahan \cite{JG} in a plaquette glass
model.
However a detailed theoretical description of this relation is lacking.

The aim of this letter is to present the first analytic calculation
of the mosaic length scale, and its comparison with the dynamical
length computed in a unified framework.
To this end we study a disordered model with Kac interactions 
\cite{FT,SF} and define length scales through properly chosen 
point-to-set correlation functions. 
We are thus able to answer some fundamental open questions, such as 
their relative order of magnitude in different temperature regimes.

It is well known that conventional static correlation functions 
do not show any signature of a large length scale in supercooled
liquids. In order to circumvent this problem, we define length scales
through ``point-to-set'' correlations \cite{BB2,MS1}.  We fix a
reference configuration $\us^{(\alpha)}=\{s^{(\alpha)}_x\}_{x\in
  \integers^d}$ drawn from the equilibrium Boltzmann measure, and
considers a second configuration $\us$ that is constrained to be close
to $\us^{(\alpha)}$ outside a sphere $\Ball_{x_0}(\ell)$ of radius
$\ell$ centered at a particular point $x_0$.  In introducing the
static correlation length $\ell_{\rm s}$, $\us$ is distributed
according to the Boltzmann law inside $\Ball_{x_0}(\ell)$ with a
boundary condition determined by $\us^{(\alpha)}$. We then define
$\ell_{\rm s}$ to be the smallest value of $\ell$ such that the
correlation between $\us^{(\alpha)}$ and $\us$ at the center of
$\Ball_{x_0}(\ell)$ (as measured through a correlation function, e.g.
$\<s^{(\alpha)}_{x_0}s_{x_0}\>_{\rm conn}$) decays below a preassigned
value $\ve$.

Consider now a particular choice of local stochastic dynamics
verifying detailed balance. Let $\tau$ be the autocorrelation time for
a local degree of freedom.  As shown in Ref.~\cite{MS2}, in a finite
range system, one has necessarily $C_1 \ell_{\rm s} \le \tau \le
\exp(C_2 \ell_{\rm s}^d)$, with $C_1$ and $C_2$ two model dependent
constants. The lower bound corresponds to a MCT-like dynamics, a good
description for $T\gg T_{\rm d}$ (up to a non-trivial scaling
exponent, i.e. $\tau\sim \ell_{\rm s}^z$ with $z\ge 1$).  The upper
bound corresponds instead to the activated behavior that should hold
below $T_{\rm d}$ (we expect $\tau\sim \exp(\ell_{\rm s}^{\psi})$ with
$\psi\le d-1$).  In a model with large-but-finite
interaction range $\gamma^{-1}$ (see below), one finds $\ell_{\rm
  s}(T,\gamma)\approx \gamma^{-1}\elb_{\rm s}(T)$, and the two types
of dynamics can be distinguished sharply (exponentially in
$1/\gamma$).

The two regimes can be bridged defining a new dynamical
length $\ell_{\rm d}$ which separates scales $\ell\gg \ell_{\rm d}$
on which non-activated relaxation is possible, from scales 
$\ell\ll \ell_{\rm d}$ on which activation is necessary. 
In order to precise this notion, consider a system initialized by setting
$\us(t=0)=\us^{(\alpha)}$ and constrained to remain close to  $\us^{(\alpha)}$ 
outside $\Ball_{x_0}(\ell)$ at all subsequent 
times, and let  $\tau(\ell)$ be the shortest time such that
$\<s_{x_0}(0)s_{x_0}(t)\>_{\rm conn}$ decays below $\ve$. 
By the above definition $\tau(\ell)=\infty$  for $\ell<\ell_{\rm s}$.
We define $\ell_{\rm d}$ by the property that $\tau(\ell)$ is
(exponentially) divergent as $\gamma\to 0$ for $\ell<\ell_{\rm d}$, and 
stays bounded for $\ell>\ell_{\rm d}$.
\begin{figure}
\includegraphics[width=8.cm]{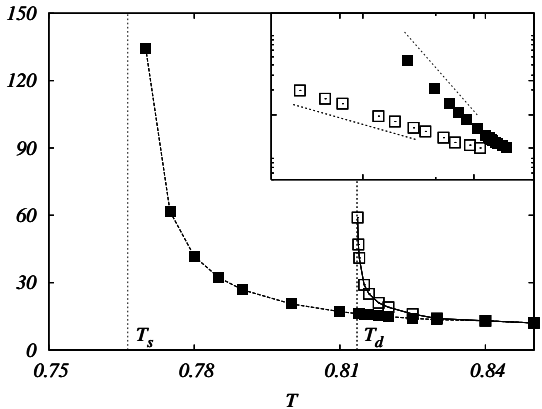}
\caption{Dynamic (empty squares) and static (filled squares) rescaled
lengths $\elb_{\rm d}$ and $\elb_{\rm s}$, for a Kac model with
two and four spin interactions. The vertical lines correspond to the 
mode coupling and thermodynamic glass transitions $T_{\rm d}\approx 0.813526$ 
and $T_{\rm s}\approx 0.766287$. 
In the inset, the rescaled lengths as functions
of $(T-T_{\rm d})$ (for $\elb_{\rm d}$) and 
$(T-T_{\rm s})$ (for $\elb_{\rm s}$). The dotted lines have slope
(respectively) $-1/4$ and $-1$.}
\vspace{-0.2cm}
\label{fig_len}
\end{figure}

In the following we shall compute $\ell_{\rm s}$ and $\ell_{\rm d}$ in
a $d$-dimensional spherical $p$-spin disordered model with Kac
interactions. Its elementary degrees of freedom are soft spins
$s_i\in\reals$, associated to the vertices of a cubic lattice of size
$2L$ (i.e. $i\in\{-L+1,\dots,L\}^d$), with periodic boundary
conditions.  Given the interaction range $\gamma^{-1}>0$ and a
non-negative rapidly decreasing function $\psi:\reals_+\to\reals_+$,
normalized by $\int\!\psi(|x|)\;\de^{d}x = 1$, we define the local
overlap of two configurations $\us^{(1)}$ and $\us^{(2)}$ as
$Q_{12}(i) = \gamma^{d}\sum_j \psi(\gamma|i-j|) \,
s^{(1)}_js^{(2)}_j$.  The \emph{local} spherical constraint is
$Q_{11}(i) = \gamma^{d}\sum_j \psi(\gamma|i-j|) \, s^{(1)}_js^{(1)}_j
= 1$.  The Hamiltonian $H(\us)$ is a quenched Gaussian random variable
of zero mean and covariance $\E[H(\us^{(1)})H(\us^{(2)})] = \sum_i
f\big(Q_{12}(i)\big)$ where $f(x)$ is a polynomial with positive
coefficients. In the following we shall use as running example the
polynomial $f(x) = \frac{1}{10}\, x^2 + x^4$ (i.e. a model with two
and four spin interactions) in $d=2$ dimensions (but the critical
behavior is independent of $d\ge 2$). While the quartic term in the
polynomial assures the wanted phenomenology at the mean-field level,
the relatively small quadratic term is introduced to have a regular 
gradient expansion of the free-energy functional (see below and \cite{SF}). 
The Kac limit is defined as $L\gg\gamma^{-1}\gg 1$ (in
other words, the limit $\gamma\to 0$ is taken \emph{after}
$L\to\infty$).  It is convenient to define the rescaled lengths
$\elb_{{\rm s}/{\rm d}}\equiv \gamma\,\ell_{{\rm s}/{\rm d}}$ that
admit a finite limit as $\gamma\to 0$.  In Fig.~\ref{fig_len} we plot
$\elb_{\rm d}$ and $\elb_{\rm s}$ as functions of temperature for our
running example, in the Kac limit.  They diverge at two distinct
temperatures $T_{\rm d}$ and $T_{\rm s}$. The dynamical length
diverges as $\elb_{\rm d}\sim (T-T_{\rm d})^{-1/4}$. This is the same
behavior as (independently) found within MCT for the length $\xi_{\rm
  d}$ \cite{chi4,arichi4}, we therefore identify $\ell_{\rm d}$ with
$\xi_{\rm d}$.  The static length behaves as $\elb_{\rm s}\sim
1/\Sigma(T)\sim (T-T_{\rm s})^{-1}$ corresponding to $\theta=d-1$.

We shall now explain how  the lengths $\ell_{\rm d}$ and 
$\ell_{\rm s}$ have been computed. 
Consider a reference configuration $\us^{(\alpha)}$, a sphere 
$\Ball_0(\ell)$ with radius $\ell$, centered at $x=0$, and let $\qb\le 1$.
We introduce the constrained Boltzmann measure $\<\,\cdot\,\>^{\alpha}_{\ell}$
by setting
\begin{eqnarray}
\<\cO\>^{\ell}_{\alpha}=\frac{1}{Z_{\alpha}}
\int_{\us}\!\cO(\us)\; 
e^{-\beta  H(\us)}\!\! \prod_{i\not\in \Ball_0(\ell)}\!\!
\delta\big(Q_{\us,\us^{(\alpha)}}(i)-\qb\big)\, ,
\end{eqnarray}
where $\int_{\us}$ denotes integration over configurations satisfying
the local spherical constraint.
For technical reasons it is preferable to take $\qb<1$:
$\us$ is only required to be close to $\us^{(\alpha)}$ outside 
$\Ball_0(\ell)$. 
The results are qualitatively independent of $\qb$ if this is large enough.
Next we define the correlation function
\begin{eqnarray}
G(\ell) \equiv \E\left\{ \E_{\us^{(\alpha)}}\left[\<Q_{\us,\us^{(\alpha)}}(0)
\>^{\alpha}_{\ell}\right]\right\}\, ,
\end{eqnarray}
where $\E_{\us^{(\alpha)}}$ and $\E$ denote (respectively) averages over
the reference configuration 
and the quenched disorder. The static length scale 
$\ell_{\rm s}$ is the smallest $\ell$ such that $G(\ell)\le \ve$.

The averages  $\E_{\us^{(\alpha)}}$ and $\E$ can be taken by
introducing replicas along the lines of \cite{SF}. Integrals over the spin
variables are then traded for and $nr\times nr$ matrix order parameter 
$q_{ab}(i)$. Next we rescale the position to define 
$\xb = \gamma\, i \in [-\Lb,\Lb]^d$, $\Lb \equiv \gamma L$, and 
write (with an abuse of notation) $q_{ab}=q_{ab}(\xb)$, to get
\begin{eqnarray}
G(\ell) =\lim_{{\tiny\begin{array}{c}n\to 0\\ r\to 1\end{array}}} 
\int q_{1,n+1}(0)\;\;e^{-\frac{1}{\gamma^d}
S[q_{ab}]}\;
\de [q_{ab}]\, .\label{eq:GlReplicas}
\end{eqnarray}
The dependency upon $\gamma$ is now completely explicit and 
the functional integral can be performed using the saddle point method.
Inspired by the solution of the mean field model \cite{FP}, we look for
a one-step replica symmetry breaking (1RSB) saddle point 
$q_{ab}^{{\rm\tiny 1RSB}}(\xb)$. This is characterized by 
three scalar functions $p(\xb)$, $q_1(\xb)$ and $q_0(\xb)$ and by 
a single Parisi 1RSB parameter $m$. While $p(\xb)$ is the local overlap 
between the reference configuration $\us^{(\alpha)}$ and the constrained one 
$\us$, $q_1(\xb)$ and $q_0(\xb)$ are the overlaps of two constrained 
configurations when they belong (respectively) to the same or to different
metastable states. Using this ansatz, one obtains 
$S[Q_{ab}]  = n\int\! \cL(\xb) \,\de^{d}\xb+O(n^2)$, where
\begin{eqnarray}
\cL & = & -\frac{\beta^2}{2}\left\{f(1)+2f(\psi*p)-(1-m)f(\psi*q_1)
\right.\nonumber\\
&&\hspace{-0.3cm}\left.
-mf(\psi*q_0)\right\}+\frac{1}{2} \frac{p^2-q_0}{1-(1-m)q_1-mq_0}
\label{eq:Lagrangian}\\
&&\hspace{-0.3cm}
-\frac{m-1}{2m}\log(1-q_1)-\frac{1}{2m}\log[1-(1-m)q_1-mq_0]\, ,
\nonumber
\end{eqnarray}
with the various fields being evaluated at position $\xb$.
Here we used the shorthand $\psi*q(\xb) = \int \!\psi(|\xb-\yb|)\,q(\yb)\,
\de^d\yb$.
The constraint enforcing $\us$ to be close to $\us^{(\alpha)}$ outside
$\Ball_0(\ell)$ is fulfilled by setting $p(\xb)=\qb$ for 
$\xb \not\in\Ball_0(\elb)$ (notice the rescaling of the radius:
$\elb = \gamma\ell$).

Using Eq.~(\ref{eq:GlReplicas}), we obtain 
$\lim_{\gamma\to 0}G(\elb/\gamma) = p(0)$, where $p(\xb)$ is evaluated
at the minimum of the action $S_0 = \int\! \cL(\xb) \,\de^{d}\xb$.
This implies that, as anticipated,  $\elb_{\rm s}(T) 
= \gamma\ell_{\rm s}(T,\gamma)$ has a finite $\gamma\to 0$ limit.
We minimize the action looking
 spherically symmetric saddle points and iterating numerically  the Euler-Lagrange
equations for (\ref{eq:Lagrangian}) until a 
stationary point is reached.
We then repeat for several initial values
of $p(\xb)$, $q_0(\xb)$ and $q_1(\xb)$ 
conditions, and compare the corresponding stationary points. 
\begin{figure}
\includegraphics[width=8.cm]{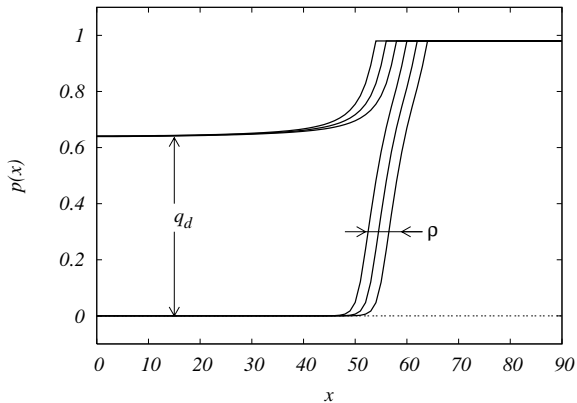}
\caption{Saddle points of the 1RSB action, cf. Eq.~(\ref{eq:Lagrangian})
for $T=0.81365$ and several values of the ball radius $\elb$.
Here we plot the outcome of the iterative solution initiated
with the $p(x)=\qb$ identically. For $\elb<\elb_{\rm d}=59$, this is
the high-overlap solution. For  $\elb>\elb_{\rm d}$
no high-overlap solution exists: the remaining dependence upon $\elb$
is essentially a shift.}
\vspace{-0.2cm}
\label{fig_prof}
\end{figure}

Since this procedure is relatively heavy
 from a computational point of
view, we simplify the Lagrangian (\ref{eq:Lagrangian}) in two ways.
First, we expand the terms of the form $f(\psi * q)(x)$ in
gradients of $q(x)$, and truncate to second order, thus
obtaining $f(q)+cf'(q)\Delta q(x)$, where $c= \frac{1}{2d}\int\!
z^2\,\psi(|z|)\,\de^d z$ (in our running example $c=1$).
Second, we discretize the model on a lattice of spacing $h$ in the
radial coordinate.
Neither of these simplifications is expected to modify the qualitative
behavior of the model, and we checked our predictions against
alternative discretizations.

In Fig.~\ref{fig_prof} we show some typical results
for $T\gtrsim T_{\rm d}$. For small $\elb$, a unique solution is
found. The overlap $p(x)$ between the probe replica and the reference one
is everywhere large: close to $\qb$ on the border, it saturates well
inside the ball to a smaller value $q_{\rm d}$ roughly independent of
the ball radius.  The two overlaps $q_0(x)$, $q_1(x)$ are equal and close
to $p(x)$.  As $\elb$ crosses some small value $\elb_0$, a new
solution appears, with $p(x)$ rapidly decaying to $0$ in the interior
of the ball, and the influence of the boundaries remaining confined to
a small distance $\rho$. This solution describes a constrained replica $\us$
well decorrelated from the reference one $\us^{(\alpha)}$, but has
initially higher value of the free energy and is therefore
thermodynamically irrelevant.  The static length $\elb_{\rm s}$ is the
smallest value of $\elb$, such that the free energy of the 
low-overlap solution becomes smaller than the free energy of the
high-overlap one. Above this length, the correlation between $\us$ and
$\us^{(\alpha)}$ in the center of the ball is small. However, the
high-overlap solution persists as a metastable state. If the
system is let evolve from the initial condition $\us^{(\alpha)}$ (and
is constrained outside the ball), it will take an exponential (in
$1/\gamma$) time to equilibrate.  The high-overlap solution disappears
at a larger radius $\elb_{\rm d}$.  For $\elb>\elb_{\rm d}$, the system
is no more trapped in a metastable state
selected by the reference configuration.
\begin{figure}
\includegraphics[width=7.5cm]{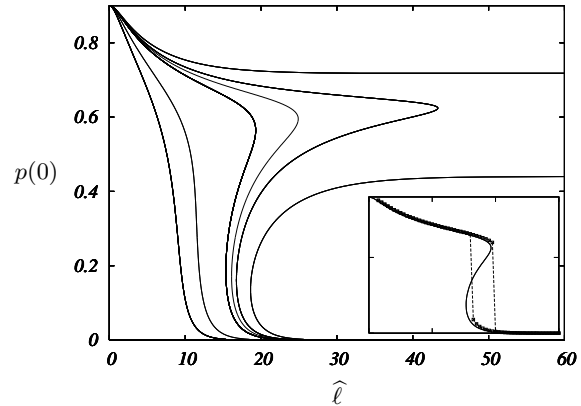}
\put(-100,-10){$\elb$}
\put(-220,75){$p(0)$}
\caption{Plot of the overlap $p(0)$ as a function of $\elb$ in
the approximated model with for  $T=1.0101, 0.9065,  0.8319,  0.8222, 
0.81510, 0.8142 > T_{\rm d}$ and $T=0.7990< T_{\rm d}$.
Inset: comparison between the approximate (analytical)
and complete (numerical) minimization of the action for $T=0.8222$.}
\vspace{-0.2cm}
\label{fig_naso}
\end{figure}

The evolution of different solutions as a function of
$\elb$ can be followed on Fig.~\ref{fig_naso}, where we plot the
overlap $p(0)$ between $\us^{(\alpha)}$ and $\us$ at the
center of the ball. For $T\gg T_{\rm d}$, $p(0)$ is a single valued function 
of $\elb$. Closer to $T_{\rm d}$, multiple branches develop,
with the coexistence region diverging as $T\downarrow T_{\rm d}$.
Beyond the high- and low-overlap solutions described
above, an intermediate unstable branch is also present.  In
order to recover the full curve (including the unstable branch) we
resorted to a simple approximation: we minimized the action under the
constraint $p(x) = q_0(x) = q_1(x)$. In this setting, and under the
``thin wall approximation'' \cite{L1}, the problem is equivalent to a
one-dimensional mechanical system, and can be solved by quadratures.
As shown in the inset of Fig.~\ref{fig_naso}, the approximate and
complete actions give results in good agreement. In particular, both
cases yield the critical behaviors $\ell_{\rm s}\sim (T-T_{\rm
s})^{-1}$ and $\ell_{\rm d}\sim (T-T_{\rm s})^{-1/4}$. Below $T_{\rm d}$ all
three branches extend to infinity, indicating that activation is
necessary on all length scales.
\begin{figure}
\includegraphics[width=8.cm]{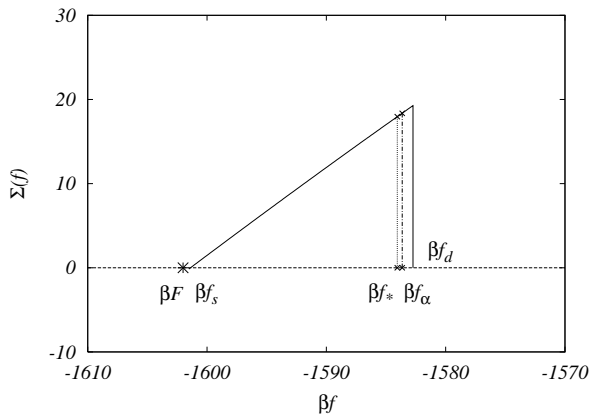}
\caption{Typical shape of the configurational entropy $\Sigma_{\elb}(f)$ for
$\elb_{\rm s}<\elb< \elb_{\rm d}$ (here $T=0.814>T_{\rm d}$, 
$\elb_{\rm s}\approx 16$, $\elb_{\rm d}\approx 41$, and $\elb= 25$). 
Inset: configurational entropy $\Sigma_{\elb}(f_*)/\elb^d$
of the dominating metastable states
as a function of $\elb$. The curve crosses the axis
at $\elb^{\1RSB}$ and ends at $\elb_{\rm d}$.}
\vspace{-0.2cm}
\label{fig_sigmaf}
\end{figure}

The complete 1RSB lagrangian (\ref{eq:Lagrangian}) allows
for a more precise description of the system. Consider $T$ close to
$T_{\rm d}$ and $\elb\in[\elb_0,\elb_d]$ where
$\elb_0$ is the minimal length for multiple solutions.  As
anticipated, on the low overlap solution, we find $q_1> q_0$.
The interpretation is, as usual, that this
solution does not describe a liquid phase, but rather a glassy one,
with many pure states.
The internal overlap of each such state is $q_1$ and the overlap
between distinct states $q_0$.
The number of metastable states is exponential in $\ell = \elb/\gamma$
for $\elb>\elb^{\1RSB}$ (the action is maximized by $m=1$)
and subexponential for $\elb<\elb^{\1RSB}$ ($m<1$). 
RSB slightly modify the free energy of the low overlap solution
for $\elb<\elb^{\1RSB}$ thus changing the value of $\elb_{\rm s}$.

Let us reconsider the thought
experiment of letting the system evolve starting from the reference
configuration $\us^{(\alpha)}$.  The system escapes the metastable
state selected by $\us^{(\alpha)}$ in a time exponential in
$1/\gamma$. After that, however, it does not equilibrate freely within
the rest of phase space, but rather moves through activation from one
state to the other, spending an exponential time in each one.

By taking the Legendre transform of the $m$-dependent free energy,
 we can compute the configurational entropy
$\Sigma_{\elb} (f)$ (i.e. the log number of metastable states as a
function of their {\it internal} free-energy $f$) \cite{remi}. In
Fig.~\ref{fig_sigmaf} we show an example of its behavior for
$\elb_{\rm s}<\elb <\elb_{\rm d}$.  An exponential number of
metastable states is present in an $\elb$-dependent free energy
interval $[f_1,f_2]$ (most of them having free energy $f_2$). The
dominating ones have free energy $f_*$ determined by maximizing
$\Sigma_{\elb}(f)-\beta f$. The free energy $F_+$ of the high overlap
state is lower than for most of other metastable states $F_+<f_2$ 
(because of the better matching with the boundaries) but higher
than the overall free energy $F_-= f_* - T\Sigma_{\elb}(f_*)$.  If
$\elb$ is decreased, the gap between $F_-$ and $F_+$ decreases.
At $\elb^{\1RSB}$ the configurational entropy $\Sigma_{\elb}(f_*)$
vanishes and $F_-=f_*=f_1$. Eventually $F_-$ and $F_+$ 
cross at $\elb=\elb_{\rm s}<\elb^{\1RSB}$. If on the other hand $\elb$
is increased, the configurational entropy increases as well, until the
1RSB solution ceases to exist.  This happens at the
dynamical length $\elb_{\rm d}$, coherently with the fact that beyond
this scale relaxation does not require activation.

For a small non-vanishing $\gamma$, the above picture should remain
roughly correct. However, activated and MCT time scales must now be
compared.  The first one corresponds a typical relaxation time
$\tau_{\rm act} \sim \exp\{\Upsilon\ell_{\rm s}^{\psi}\}$, while the
second gives $\tau_{{\rm{\footnotesize MCT}}} \sim \ell_{\rm d}^z$.
Therefore, MCT behavior is faster and dominates until very close to
$T_{\rm d}$. The divergence is then cut-off when $\ell_{\rm d}(T)
\approx \exp\{\Upsilon\ell_{\rm s}(T_{\rm d})^{\psi}/z\}$, and
dynamics becomes activated at lower temperatures.  During this
cross-over, the physical dynamical length $\xi_d$ characterizing the
size of cooperatively rearranging regions crosses over from $\xi_d
\approx \ell_{\rm d}$, to $\xi_d \approx \ell_{\rm s}$.

Summarizing we implemented the mean-field theory of mosaic state in a
microscopic model. This allowed to establish relations between mosaic and 
MCT dynamical lengths which had not been predicted from phenomenological
considerations. 
The dynamical length scale is defined by the property that 
dynamics is dominated by activation at shorter scales. 
It can further be  identified with the one appearing in 
the MCT divergence of four point susceptibilities.
Below mosaic length the system is instead rigid for thermodynamic reasons. 
The two happen to be close above the MCT critical temperature and widely
separated between this and the glass transition, corresponding
to the crossover between relaxational and activated dynamics.

{\bf Acknowledgments} 
We thank the participants of the programme
``Principles of the Dynamics of Non-Equilibrium System'' 9 January -
30 June 2006, Isaac Newton Institute, Cambridge and in particular D.
Mukamel for important discussions.
This work was supported in part by the
EU under contract
``HPRN-CT-2002-00319 STIPCO''.

\end{document}